# Exact Description of Black Holes on Branes

ROBERTO EMPARAN[a], GARY T. HOROWITZ[b], ROBERT C. MYERS[c]

[a] *Departamento de Física Teórica, Universidad del País Vasco, Apdo. 644, E-48080 Bilbao, Spain*

[b] *Physics Department, University of California, Santa Barbara, CA 93106 USA*

[c] *Institute for Theoretical Physics, University of California, Santa Barbara, CA 93106 USA*

[c] *Department of Physics, McGill University, Montréal, QC, H3A 2T8, Canada*

[a]wtpemgar@lg.ehu.es, [b]gary@cosmic.physics.ucsb.edu, [c]rcm@hep.physics.mcgill.ca

## Abstract

To test recent ideas about lower dimensional gravity bound to a brane, we construct exact solutions describing black holes on two-branes in four dimensions. We find that $2 + 1$ gravity is indeed recovered at large distances along the brane, although there are significant deviations at smaller scales. Large black holes appear as flattened "pancakes", with a relatively small extent off the brane. Black hole thermodynamics is discussed from both the standpoint of the brane and the bulk. We comment on the analogous black holes bound to higher dimensional branes.

November, 1999

# 1. Introduction

Randall and Sundrum have recently argued that four-dimensional gravity can naturally arise at long distances on a three-brane embedded in five dimensions [1]. They started by taking two copies of (a part of) five-dimensional anti-de Sitter space ($AdS_5$) and gluing them together along a boundary which could be interpreted as the three-brane [2]. They then studied linearized fluctuations about this background. There is a zero mode which looks like a four-dimensional graviton bound to the brane, together with a continuum of massive Kaluza-Klein modes. They argued that the zero mode produces the standard $1/r$ gravitational potential along the brane, and the Kaluza-Klein modes give rise to negligible corrections of order $\ell^2/r^3$ where $\ell$ is the radius of curvature of $AdS_5$.

This intriguing alternative to compactification has attracted considerable interest. (For earlier work on related ideas, see [3] and references therein.) However, there is a potential problem[1] The background solution contains $AdS_5$ horizons which are infinitely far from the brane in spacelike directions, but can be reached by observers in finite proper time (just like the horizons of extreme charged black holes). The linearized modes studied in [1] turn out to be singular on these horizons [5]. Since there is no reason for physics on the brane to produce curvature singularities elsewhere in spacetime, there is some concern that the right boundary conditions have not been imposed. Alternatively, the singularities may just be an artifact of the expansion in plane waves. It is possible that when localized sources and wave packets are considered, the singularities will be removed.

To test this, we consider the analogous question in one lower dimension. In this case we are not limited to a linearized analysis about an $AdS_4$ background. Instead, one can construct exact four-dimensional solutions describing localized black holes bound to a two-brane. To obtain them, notice that a black hole on a brane in $AdS$ is accelerating. There is a solution to Einstein's equation that describes accelerating black holes, called the C-metric [6]. This solution can be extended to include a cosmological constant [7]. Of course, one needs a mechanism to accelerate these black holes, so these solutions usually contain conical singularities along the axis from the black hole to infinity. These singularities can be interpreted as cosmic strings pulling on the black holes. To obtain our solutions, we find an appropriate three-surface $\Sigma$ such that there are no conical singularities on one side of $\Sigma$. We then glue two copies of this part of the spacetime together along $\Sigma$, and verify that the induced stress tensor on $\Sigma$ is proportional to the induced metric. This ensures that $\Sigma$ can be interpreted as a two-brane. The net result is a solution without conical singularities, with a smooth $AdS$ horizon, and containing a black hole bound to the Planck two-brane.

---

[1] We will be interested in the case where the fifth direction is infinite. If one makes it finite by introducing a second brane with negative tension [4], this potential problem cannot arise.

The key question now is whether lower dimensional gravity is recovered on the brane. Three-dimensional gravity is rather special since spacetime is locally flat outside the matter, and the total mass is reflected in an asymptotic deficit angle [8,9]. Nevertheless, we do recover $2+1$ gravity at large distances along the brane. Since the mass affects the leading order $2+1$ metric rather than a subleading correction (as in higher dimensions), it is perhaps even more remarkable that this is recovered from a four-dimensional solution. We believe this is strong support for the Randall-Sundrum scenario.

Deviations from $2+1$ gravity arise at order $1/r$, reflecting the four-dimensional nature of the black hole. For large black holes, these deviations can be significant at large scales. Large black holes look like flattened pancakes, since they are much larger in the directions along the brane than in directions off the brane. There is a maximum mass in $2+1$ gravity corresponding to the fact that the deficit angle cannot be greater than $2\pi$ [10]. We will see that the area of the black hole grows without bound as the mass approaches this limit. Thus arbitrarily large black holes can be produced from finite mass! This unusual property appears to be a special feature of black holes on two-branes, and we do not think it will continue in the more realistic case of three-branes. However, our results suggest that in the higher dimensional solution, corrections to the $1/r$ potential on a three-brane should start at order $1/r^2$, not $1/r^3$ as suggested in [1]. This would just reflect the five-dimensional character of the black hole.

We also study the thermodynamics of the black holes we construct. Since the black hole temperature is constant over the horizon, it is the same in the bulk and on the brane. The entropy is one-quarter of the area of the full four-dimensional black hole. Since the four-dimensional solution is not asymptotically AdS in the usual sense, one cannot define a four-dimensional total mass in the usual way. Nevertheless, one can integrate the thermodynamic relation $dM = TdS$ to obtain an effective mass for the black hole. This turns out to agree exactly with the $2+1$ mass obtained from the asymptotic geometry along the brane. Even though the mass and temperature both agree when computed in the bulk and on the brane, the entropy is *not* one-quarter of the circumference of the horizon on the brane. This is hardly surprising since there are no pure $2+1$ black holes (in the absence of a cosmological constant). However, we find that in the limit of large black holes, the entropy reduces to one-quarter of the circumference, up to small corrections. We will argue that this result holds in the higher dimensional case as well, due to the AdS geometry.

The outline of this paper is the following. In the next section, we extend the Randall-Sundrum construction to general dimensions, and review the potential problem with singularities. In section 3, we construct the solutions describing black holes on a two-brane, show that $2 + 1$ gravity is recovered and investigate the thermodynamic properties of the black hole. The last section contains a discussion of the implications of these results for the more realistic higher dimensional solution.

## 2. Randall-Sundrum scenario in $n + 1$ dimensions

While the original analysis of Randall and Sundrum [1] was made in five dimensions, it is easily extended to an arbitrary number of spacetime dimensions. Furthermore, their treatment of the linearized graviton bound to the brane can be extended to include the full nonlinear equations. To see this, we start with the following metric in $n + 1$ dimensions

$$ds^2 = \hat{g}_{AB} dx^A dx^B = \frac{\ell^2}{z^2} \left[ g_{\mu\nu}(x) dx^\mu dx^\nu + dz^2 \right] \ . \tag{2.1}$$

If the $n$-dimensional metric $g_{\mu\nu}$ corresponds to flat space, $i.e.$, $g_{\mu\nu} = \eta_{\mu\nu}$, the full metric (2.1) describes $AdS_{n+1}$ in Poincare coordinates. Conformal infinity is reached at $z = 0$, while $z = \infty$ corresponds to the usual AdS horizon. One may note [11][12], however, that eq. (2.1) continues to satisfy Einstein's equations with a negative cosmological constant, $i.e.$, $R_{AB}(\hat{g}) = -(n/\ell^2) \hat{g}_{AB}$, provided $g_{\mu\nu}$ is Ricci-flat, $i.e.$, if $g_{\mu\nu}$ satisfies the vacuum Einstein equations in $n$ dimensions: $\mathcal{R}_{\mu\nu}(g) = 0$. The curvature in the modified metric now satisfies

$$R_{ABCD} R^{ABCD} = \frac{2n(n+1)}{\ell^4} + \frac{z^4}{\ell^4} \mathcal{R}_{\mu\nu r\sigma} \mathcal{R}^{\mu\nu r\sigma} \ . \tag{2.2}$$

Two points are noteworthy here: First, a curvature singularity generically appears on the 'horizon' at $z = \infty$ [11][12]. Second, the asymptotic curvature falls off much more slowly as $z \to 0$ than is usually considered for perturbations of $AdS_{n+1}$ — see $e.g.$, [13]. As a result, these metrics are not asymptotically $AdS$ in the usual sense. If one started with a metric $g_{\mu\nu}$, it would require infinite energy to change it.

The Randall-Sundrum geometry [1] is constructed by cutting off the asymptotic (small $z$) region of eq. (2.1) at the surface $z = \ell$, and completing the space by gluing onto this surface a mirror copy of the large $z$ geometry. Translating $\hat{z} = z - \ell$ so the brane resides at $\hat{z} = 0$, we may write the resulting metric as

$$ds^2 = \frac{\ell^2}{(\ell + |\hat{z}|)^2} \left[ g_{\mu\nu}(x) dx^\mu dx^\nu + d\hat{z}^2 \right] \ . \tag{2.3}$$

This metric is continuous but not differentiable. Using the standard Israel junction conditions [14] (see also [15]), the discontinuity in the extrinsic curvature is interpreted in terms of a $\delta$-function source of stress-energy at $z = \ell$. In the present case, the extrinsic curvature of the $z = \ell$ surface reduces to

$$K_{\mu\nu} = \frac{1}{2} n^\sigma \partial_\sigma \hat{g}_{\mu\nu} = \frac{1}{\ell} g_{\mu\nu} \tag{2.4}$$

where $n^\sigma \partial_\sigma = -\partial_z$ is the outward directed unit normal vector. Then defining the discontinuity in the extrinsic curvature across the gluing surface, $\gamma_{\mu\nu} \equiv K_{\mu\nu}^+ - K_{\mu\nu}^- = 2K_{\mu\nu}^+$, the surface stress-tensor becomes

$$S_{\mu\nu} = \frac{1}{8\pi G_{n+1}} \left( \gamma_{\mu\nu} - g_{\mu\nu} \gamma^\sigma{}_\sigma \right) = -\frac{n-1}{4\pi G_{n+1} \ell} g_{\mu\nu} \ . \tag{2.5}$$

The source of stress-tensor can then be interpreted as a thin relativistic $(n{-}1)$-brane with tension $T_{n-1} = (n-1)/(4\pi G_{n+1}\ell)$, where $G_{n+1}$ is the Newton's constant in $n+1$ dimensions.

The interesting feature of the Randall-Sundrum geometry is that perturbations of the $n$-dimensional metric $g_{\mu\nu}$ are now normalizable modes peaked at $z = \ell$ or $\hat{z} = 0$. In the analysis of [1], those perturbations satisfying $\mathcal{R}_{\mu\nu}(g) = 0$ correspond to the zero-mode which yields $n$-dimensional gravity on the brane. As in [1] by considering metrics of the form (2.3), one can derive an effective action for the perturbations of $g_{\mu\nu}(x)$. One starts with the $(n+1)$-dimensional gravity action

$$I = \frac{1}{16\pi G_{n+1}} \int d^n x \, dz \, \sqrt{-\hat{g}} \left( R + \frac{n(n-1)}{\ell^2} \right) - \int d^n x \sqrt{-g_{brane}} T_{n-1} \tag{2.6}$$

which includes the contributions of a negative cosmological constant $\Lambda = -n(n-1)/2\ell^2$ in the bulk and the Planck brane with tension $T_n$. The induced metric on the brane is precisely $g_{\mu\nu}$. Now using $R = (z^2/\ell^2)\,\mathcal{R}(g) + \dots$ for eq. (2.1), one integrates over $z$ to obtain the effective $n$-dimensional action

$$\begin{aligned} I' &= \frac{2}{16\pi G_{n+1}} \int d^n x \, \sqrt{-g} \int_\ell^\infty dz \left( \frac{\ell}{z} \right)^{n+1} \left[ \left( \frac{z}{\ell} \right)^2 \mathcal{R}(g) + \dots \right] \\ &= \frac{1}{16\pi G_n} \int d^n x \, \sqrt{-g} \, \mathcal{R}(g) \end{aligned} \tag{2.7}$$

where the brane tension (2.5) has been tuned to cancel cosmological constant contributions in this action.[2] Hence the effective action governing the metric zero-mode is precisely the $n$-dimensional Einstein action where the effective Newton's constant in $n$ dimensions is given by

$$G_n = \frac{n-2}{2\ell} G_{n+1} . \tag{2.8}$$

If one chooses $g_{\mu\nu} = \eta_{\mu\nu}$, the remaining metric perturbations, *i.e.*, the Kaluza-Klein modes, of the Randall-Sundrum geometry fall into a continuum of massive representations of the $n$-dimensional Lorentz group [1]. After gauge fixing appropriately, one finds in separation of variables that $\nabla_g^2 \, \delta\hat{g}_{AB} = m^2 \, \delta\hat{g}_{AB}$. Then the radial profile for these modes satisfies a Schrödinger-like equation

$$\left(-\partial_{\hat{z}}^2 + V(\hat{z})\right)\psi(\hat{z}) = m^2 \psi(\hat{z}) \tag{2.9}$$

with a so-called 'volcano' potential

$$V(\hat{z}) = \frac{n^2-1}{4(|\hat{z}|+\ell)^2} - \frac{n-1}{\ell}\delta(\hat{z}) . \tag{2.10}$$

Here, the metric perturbations satisfy $\delta\hat{g}_{\mu\nu} \propto (|\hat{z}|+\ell)^{(n-5)/2}\psi(\hat{z})$, and one can easily confirm that $\psi \propto (|\hat{z}|+\ell)^{-(n-1)/2}$ provides the desired zero-mode bound-state.

From (2.9) and (2.10), we see that the Kaluza-Klein modes with $0 < m^2 < (n^2-1)/4\ell^2$ are suppressed at the brane by the wings of the volcano potential. One can also interpret this as the usual asymptotic suppression by the geometric potential barrier arising in $AdS_{n+1}$. This suppression is sufficient to ensure that at least on distance scales greater than $\ell$, gravity on the brane is actually $n$-dimensional Einstein gravity up to small power-law-suppressed corrections [1].

As illustrated in eq. (2.2), there is still a potential problem with the zero mode in the Randall-Sundrum geometry. That is, generically one finds a curvature singularity as $z \to \infty$. Intuitively, however, one expects that masses confined to the brane will generate localized gravitational fields, and so these singularities should not be physically relevant.

---

[2] If one does not tune the cancellation of the cosmological constant and the brane tension in the effective action (2.7), one generates an effective cosmological constant for the brane gravity [16]. The resulting geometries can be regarded as extending the above construction to other interesting slicings of $AdS_{n+1}$ [11]. We will construct black holes bound to such cosmological branes elsewhere [17].

Thus despite the suppressed coupling of the Kaluza-Klein modes, their contributions must be significant far from the brane and ensure that the total gravitational field decays to zero more rapidly than the zero-mode profile $\ell^2/z^2$. One approach to investigate this potential problem is to construct explicit solutions of the $(n+1)$-dimensional Einstein equations when a massive source is added on the Planck brane. In the next section, we will construct such a solution for the case of $n = 3$.

The case of $n = 3$ is special in that for three-dimensional gravity $\mathcal{R}_{\mu\nu} = 0$ in fact implies that $\mathcal{R}_{\mu\nu r\sigma} = 0$. The effect of a massive source on the three-dimensional geometry is to generate a deficit angle and hence outside of the sources where the curvature vanishes, the geometry is conical [8,9]. There are no black hole solutions for three-dimensional Einstein gravity (with vanishing cosmological constant). Rather, point masses generate conical singularities. Considering just the zero mode in this context would correspond to setting $g_{\mu\nu}$ in (2.1) to be the metric of a cone. The resulting solution has conical singularities extending throughout the spacetime, not just on the brane. Our results in the next section show that such solutions are not physically relevant. Rather, gravitational collapse of sources on the brane will produce four-dimensional black holes, whose singularities lie on the brane, and are surrounded by horizons that extend only slightly off the brane. Nevertheless, a deficit angle is still present at large distances along the brane.

Note that, in the above discussion, the brane was placed at $z = \ell$ rather than some general $z = \alpha\ell$. However this is just a coordinate choice, and can be changed by rescaling $z$ in (2.1). It may seem a little surprising that there is no physics in the location of the Planck brane, since as $z \to \infty$ one approaches the horizon, and one might have expected the acceleration required to stay at constant $z$ to increase. However, one can easily verify that *e.g.*, if $g_{\mu\nu} = \eta_{\mu\nu}$ in eq. (2.1), the magnitude of the acceleration of the worldlines at constant $z$ (and $x^i$) is always $1/\ell$, and does not diverge as $z \to \infty$. Similarly, one can easily confirm that $K_{\mu\nu} = (1/\ell)\,\tilde{g}_{\mu\nu}$ where $\tilde{g}_{\mu\nu}$ is the induced metric on the brane independent of the choice for the location of the brane.

## 3. Black holes on a two-brane

### 3.1. Constructing the solution

In this section we will construct a solution describing a black hole on the Planck brane in the 3+1 analog of the Randall-Sundrum scenario. We begin with the following solution of Einstein's equation with a negative cosmological constant:

$$ds^2 = \frac{\ell^2}{(x-y)^2} \left[ -(y^2 + 2\mu y^3)dt^2 + \frac{dy^2}{(y^2 + 2\mu y^3)} + \frac{dx^2}{G(x)} + G(x)d\varphi^2 \right] \tag{3.1}$$

where

$$G(x) = 1 - x^2 - 2\mu x^3 \tag{3.2}$$

One can verify that this metric satisfies $R_{AB} = -(3/\ell^2)\hat{g}_{AB}$. The metric (3.1) is a special case of a more general solution called the AdS C-metric [7]. It is clearly invariant under translations of $t$ and $\varphi$.

Since the metric is written in terms of an unusual coordinate system, its physical interpretation is not immediately clear. Roughly speaking, $y$ is like a radial variable, and $x$ is analogous to $\cos\theta$. To begin to understand this solution, let us consider the range of these coordinates. If $0 < \mu < 1/3\sqrt{3}$, then the cubic $G(x)$ has three real roots $x_0 < x_1 < 0 < x_2$. We will discuss this case first. Since we need $G(x) \geq 0$ in order for the metric to have Lorentz signature, we restrict $x$ to lie in the range $x_1 < x < x_2$. The factor $(x-y)^{-2}$ in front of the metric implies that $y = x$ is infinitely far away from points with $y \neq x$. So we restrict $y$ to satisfy $-\infty < y < x$. It turns out that there is a curvature singularity at $y = -\infty$. This singularity is not visible to all observers since there is a black hole horizon at $y = y_0 = -1/2\mu$. There is a second horizon at $y = 0$ which is degenerate, i.e., has zero Hawking temperature (see Fig. 1), which we will see corresponds to the AdS horizon. Since $G(x)$ vanishes at $x = x_1, x_2$ these directions correspond to the axis of rotation (like $\theta = 0, \pi$). To avoid a conical singularity at $x = x_2$, we must choose $\varphi$ to have period

$$\Delta\varphi = \frac{4\pi}{|G'(x_2)|} \tag{3.3}$$

However, once we have done this, we are no longer free to adjust the period of $\varphi$ at $x = x_1$. In general, there is a conical singularity along this axis corresponding to a deficit angle

$$\delta = \frac{4\pi}{G'(x_1)} - \Delta\varphi \tag{3.4}$$

One can think of this as a cosmic string extending from the black hole out to asymptotic infinity. The surfaces of constant $y$ are topological spheres for $y < x_1$. But if $y \geq x_1$ is held constant, then $x$ has only one axis at $x = x_2$ (since it must remain greater than $y$). The constant $y$ surfaces become topologically $R^2$.

If $\mu > 1/3\sqrt{3}$, the two roots $x_0$ and $x_1$ become complex. In this case, $G(x) > 0$ for all $x < x_2$, hence the allowed range of $x$ becomes $y \leq x \leq x_2$. In this case, one can still reach $x = y$ at $y = y_0 = -1/2\mu$ and so the event horizon reaches out to asymptotic infinity. One can think that the cosmic string above is replaced by a semi-infinite black string extending out to asymptotic infinity. In this regime for all $y$, $x$ has only one axis and the constant $y$ surfaces are topologically $R^2$.

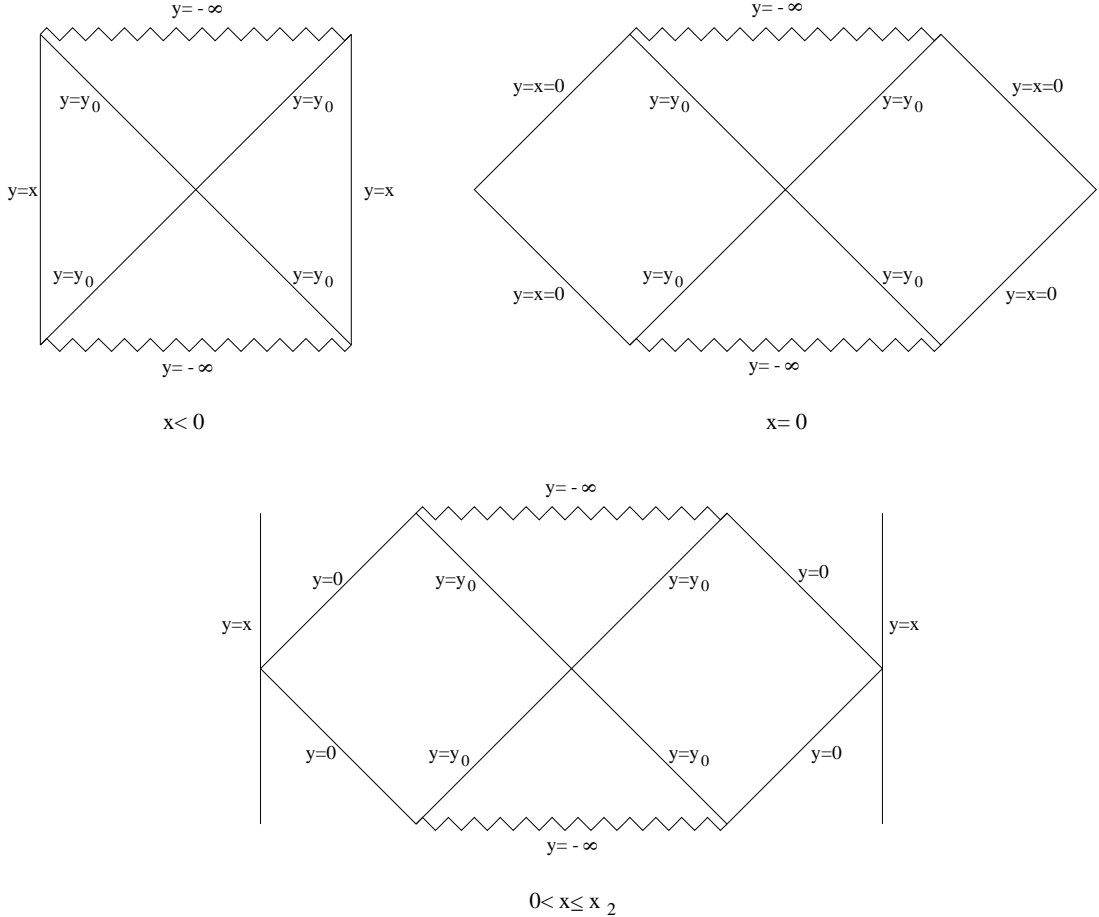

Fig. 1: Causal structure of (3.1) on sections of constant $x$. For $x \neq 0$ the sections are asymptotically $AdS_3$. The section at $x = 0$ is like the radial geometry of a four-dimensional Schwarzschild black hole. The black hole horizons are at $y = y_0 = -1/2\mu$. The horizon $y = 0$ is analogous to the AdS horizon at $z = \infty$. Points in these diagrams are circles parametrized by $\varphi$, which become of zero size at $x = x_2$, and possibly at $x = x_1 < 0$.

To get oriented, let us set $\mu = 0$. Then $-1 \leq x \leq 1$, $\Delta\varphi = 2\pi$, and the black hole horizon disappears (together with its curvature singularity inside). One can check that the Weyl tensor vanishes, so (3.1) must be $AdS_4$ in disguise. To obtain a more familiar form

of the metric, set

$$z = \ell \left( 1 - \frac{x}{y} \right), \qquad r = -\frac{\ell \sqrt{1-x^2}}{y}, \qquad \hat{t} = \ell t. \tag{3.5}$$

Then (3.1) with $\mu = 0$ becomes (compare to eq. (2.1))

$$ds^2 = \frac{\ell^2}{z^2}(-d\hat{t}^2 + dr^2 + r^2 d\varphi^2 + dz^2) \tag{3.6}$$

Notice that $y = -\infty$ becomes the nonsingular worldline $z = \ell, r = 0$. The region $x = y$ corresponds to the standard boundary at infinity $z = 0$, and $y = 0^-$ corresponds to the usual AdS horizon in Poincare coordinates $z = \infty$. The coordinate transformation (3.5) only holds for $y < 0$, but the metric (3.1) with $\mu = 0$ holds for all $y < x$ with $-1 \leq x \leq 1$ (with only the usual coordinate singularity associated with the AdS horizon). This covers two Poincare patches, one on each side of the horizon $y = 0$. For $y > 0$, the Poincare coordinates can be obtained by simply changing the overall signs in the definition of $z$ and $r$ in (3.5). We only scaled the time coordinate in eq. (3.5),[3] so the time translation symmetry in (3.1) reduces to that for the Poincare time when $\mu = 0$.

Now consider the solution with $\mu > 0$. This introduces a black hole with horizon at $y = y_0 = -1/2\mu$. For small $\mu$, the causal structure outside the black hole is shown in Fig. 2. To obtain a solution describing a black hole on a two-brane, we repeat the construction of section 2. Hence first, we must find a timelike three-surface $\Sigma$ whose extrinsic curvature is proportional to its intrinsic metric. On one side of $\Sigma$, there should be no conical singularity. One can then take two copies of this side of the spacetime, and glue them together along $\Sigma$. The resulting spacetime will be free of conical singularities.

It turns out to be remarkably easy to find a suitable $\Sigma$: We can take $\Sigma$ to be the surface $x = 0$. There are no conical singularities in the region $x \geq 0$. The extrinsic curvature $K_{\mu\nu}$ of $\Sigma$ can be calculated as in (2.4). The unit outward normal to the surfaces of constant $x$ is $-(\frac{x-y}{\ell})\sqrt{G(x)}\partial_x$. Then

$$K_{\mu\nu} = -\frac{(x-y)\sqrt{G(x)}}{2\ell} \frac{\partial \hat{g}_{\mu\nu}}{\partial x} \tag{3.7}$$

where the derivative is evaluated at $x = 0$. Since $G'(0) = 0$, the only contribution to the derivative comes from the overall factor of $(x-y)^{-2}$ in front of the metric. Using $G(0) = 1$, one finds $K_{\mu\nu} = (1/\ell)g_{\mu\nu}$ where $g_{\mu\nu}$ be the induced metric at $x = 0$. Thus our final spacetime is obtained by taking two copies of the region $0 \le x \le x_2$ in the metric (3.1) and gluing them along $x = 0$. The extrinsic curvature is precisely as in eq. (2.4), and so the surface $x = 0$ is interpreted as a thin relativistic two-brane with tension $T_2 = 1/(2\pi G_4 \ell)$.

---

[3] Note that all of the coordinates in the metric (3.1) are dimensionless, while the coordinates $\hat{t}$, $r$ and $z$ have the expected dimension of length.

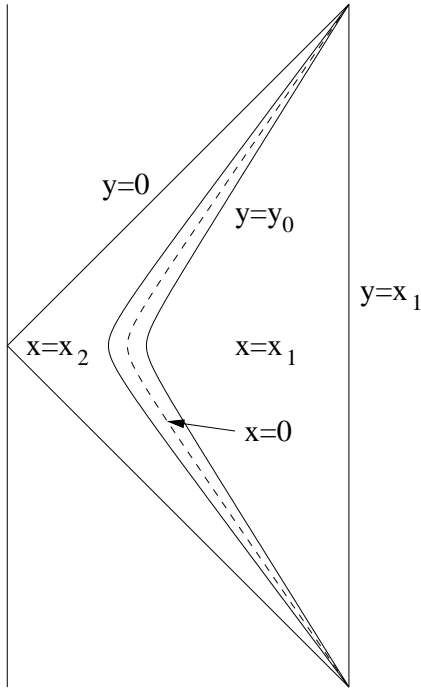

Fig. 2: (Pseudo-)Penrose diagram showing the structure of the metric (3.1) in the region outside a small black hole (small $\mu$). The black hole trajectory is indicated as the region enclosed between two solid lines that correspond to $y = y_0 = -1/2\mu$. The two-brane at $x = 0$ ($z = \ell$) cuts across the black hole horizon, and is indicated by a dashed line. In the Randall-Sundrum model the region to the left of this surface ($x_2 < x < 0$) is glued along $x = 0$ onto a copy of itself.

Having constructed the desired exact solution, we can now check the main claim of [1] that lower dimensional gravity is recovered at large distances on the brane. Let us again introduce the Poincare coordinates (3.5). The two-brane at $x = 0$ is at $z = \ell$. The black hole horizon corresponds to $z = \ell[1 - (x/y_0)] = \ell(1 + 2\mu x)$. So the black hole is centered on the brane $z = \ell$ but extends off of it to $z_{max} = \ell(1 + 2\mu x_2) < \ell(1 + 2\mu)$ since $x_2$ cannot be greater than one. In contrast, the curvature singularity at $y = -\infty$ is located only on the brane $z = \ell$ at $r = 0$. The metric on the brane is

$$ds^2 = -\left(1 - \frac{2\mu\ell}{r}\right)d\hat{t}^2 + \left(1 - \frac{2\mu\ell}{r}\right)^{-1}dr^2 + r^2 d\varphi^2 \tag{3.8}$$

As noted above, gravity in $2 + 1$ dimensions (without a cosmological constant) is rather trivial since the field equations require spacetime to be locally flat outside the matter. The effect of the matter is to curve space into a cone. The metric on the brane (3.8) is clearly not a solution of the three-dimensional Einstein equations everywhere. However, for $r \gg \mu\ell$, the geometry is approximately locally flat, and the periodicity of $\varphi$ is *not* $2\pi$. Since $\varphi$ is not affected by the change to Poincare coordinates, its periodicity is still given by (3.3) to ensure that the axis $x = x_2$ is smooth. Thus, the geometry on the brane is indeed asymptotically a cone as expected for three-dimensional gravity. However, there are significant deviations from $\mathcal{R}_{\mu\nu}(g) = 0$ for $r = O(\mu\ell)$. Note that for large $\mu$, these deviations extend out to very large distances. This is an artifact of 2+1 dimensions, where there are no deviations from flatness to compete with the deviations arising from the higher dimensions. Note that the $t$-$r$ part of this metric (3.8) coincides precisely with that of the Schwarzschild geometry.

As one moves away from the brane, the effects of the black hole diminish. If we stay on the axis, $x = x_2$, it is clear from (3.5) that large $z$ corresponds to small $y$ via $y \approx -x_2\ell/z$. The ratio of the $\mu$ term in the metric to the leading term is $2\mu y = -2\mu\ell x_2/z$. So the effects of the black hole fall off like $1/z$ for large $z$, as expected for a four-dimensional black hole. Of course, the full metric $\hat{g}_{\mu\nu}$ still contains an additional overall factor of $(\ell/z)^2$, as well.

It is not just the two-brane with the black hole at $z = \ell$ that has a deficit angle asymptotically. We now show that any test brane at constant $z$ will also have the same deficit angle.[4] So $2 + 1$ gravity holds asymptotically at all values of $z$! To see this, note from (3.5) that on a surface of constant $z = \ell + \Delta z$, $x\ell = -y\Delta z$. From the definition of $r$, $r^2y^2 = \ell^2(1 - x^2) = \ell^2 - y^2\Delta z^2$ which implies

$$y = -\frac{\ell}{[r^2 + \Delta z^2]^{1/2}} \tag{3.9}$$

From these equations, it is clear that at large $r$ for fixed $\Delta z$, both $y$ and $x$ approach zero. Hence, from the metric (3.1), the $\mu$ dependent terms become negligible, so the metric approaches (3.6). However, the periodicity of $\varphi$ is unchanged and so there will be the same deficit angle on each of these test branes.

### 3.2. Limiting cases

We now consider in more detail the limiting cases $0 < \mu \ll 1$, and $\mu \gg 1$. First we

---

[4] Note that choosing to consider a test brane on surfaces of fixed $z$ is an arbitrary choice, at least for small $r$.

assume $\mu$ is small and work to first order in this quantity. The roots of $G(x)$ are shifted by $\mu$

$$x_1 \simeq -1 - \mu, \quad x_2 \simeq 1 - \mu \tag{3.10}$$

and (3.3) implies

$$\Delta\varphi \simeq 2\pi(1 - 2\mu) \tag{3.11}$$

In the full AdS-C metric (3.1), the black hole horizon at $y = y_0 \equiv -1/2\mu$ has area

$$\mathcal{A} = \ell^2 \Delta\varphi \int_{x_1}^{x_2} \frac{dx}{(x - y_0)^2} \simeq 16\pi\mu^2\ell^2 \tag{3.12}$$

which is exactly what we would expect for a Schwarzschild black hole of mass $\mu\ell/G_4$. From (3.4), the conical singularity at $x = x_1$ corresponds to a string with tension $\delta/8\pi G_4 = \mu/G_4$. So before we introduce the brane, the full metric (3.1) describes a black hole of mass $\mu\ell/G_4$ and acceleration $1/\ell$ provided by a string.

In $2+1$ gravity, the asymptotic deficit angle is related to the total mass by $2\pi - \Delta\varphi = 8\pi G_3 M_3$, where $G_3$ is the three-dimensional Newton's constant — see, for example, [10]. We will show in section 3.3 that $M_3$ is equal to the four-dimensional mass $M_4$ of the black hole for all $\mu$. For small $\mu$, $G_4 M_4$ is just $\mu\ell$ (as suggested by the horizon area). To compare with $M_3$ we must relate $G_3$ to the four-dimensional Newton's constant $G_4$ as in eq. (2.8). For $n = 3$, one has[5]

$$G_3 = \frac{G_4}{2\ell} \tag{3.13}$$

From (3.11), the deficit angle is $8\pi G_3 M_3 = 2\pi - \Delta\varphi = 4\pi\mu$ which implies $M_3 = \mu\ell/G_4 = M_4$.

After we introduce the brane, the area of the event horizon is given by an integral like (3.12), but instead of integrating from $x_1$ to $x_2$, one has twice the integral from 0 to $x_2$. It is easy to see that when $\mu \ll 1$ the answer is essentially unchanged. The event horizon extends off the brane $z_{max} - \ell \simeq 2\mu\ell$, which for $\mu \ll 1$, gives a good approximation to the proper distance which the horizon extends off the brane. Hence the geometry in the neighborhood of these small black holes is essentially the same as that for spherical four-dimensional Schwarzschild black holes.

We now investigate what happens when $\mu \gg 1$. There is an apparent puzzle in that there is an upper limit on the mass in $2 + 1$ gravity coming from the fact that the deficit angle cannot exceed $2\pi$. The latter implies $M_{3,max} = 1/(4G_3) = \ell/(2G_4)$. This would correspond to a four-dimensional mass whose Schwarzschild radius is of order the AdS radius. (When $\mu \gg 1$, the four-dimensional mass is no longer proportional to $\mu$.) Nevertheless, we now show that there is no limit to the size of the black hole that can be produced on the brane.

---

[5] Since the mass is determined from the metric at large $r$ which approaches (3.6), it is appropriate to use the analysis of section 2 to relate $G_3$ to $G_4$.

When $\mu \gg 1$, the function $G(x)$ in (3.2) has only one real root at approximately

$$x_2 = \frac{1}{(2\mu)^{1/3}} \tag{3.14}$$

As we are in the regime $\mu > 1/3\sqrt{3}$ for the original solution without the brane (3.1), $x$ now has the range $y < x \leq x_2$, and the surfaces of constant $y$ are all topologically $R^2$. Since there is only one axis, there is no conical singularity, provided we identify $\varphi$ with period (3.3)

$$\Delta\varphi = \frac{4\pi}{3(2\mu)^{1/3}} \tag{3.15}$$

The event horizon at $y = y_0 = -1/(2\mu)$ has infinite area, and as noted above corresponds to a semi-infinite black string extending to infinity. Note that as $x$ can still approach $y$ as $y \to -\infty$, the curvature singularity now also 'extends out' to asymptotic infinity.

We now make the same construction as before, slicing this solution along the surface $x = 0$. The above argument about the extrinsic curvature still applies and shows that we can match two copies of the region $0 \leq x \leq x_2$ along the surface $x = 0$. Since $\Delta\varphi \to 0$ as $\mu \to \infty$, $M_3$ approaches its maximum value in this limit. The area of the event horizon is now finite

$$\mathcal{A} = 2\ell^2 \Delta\varphi \int_0^{x_2} \frac{dx}{(x - y_0)^2} \simeq \frac{8\pi\ell^2}{3} (2\mu)^{2/3} \tag{3.16}$$

and becomes arbitrarily large as $\mu \to \infty$. In the Poincare coordinates (3.5), the metric on the brane is still given by (3.8). So the horizon is located at $r = 2\mu\ell$, and has proper circumference

$$\mathcal{C} = 2\mu\ell\Delta\varphi = \frac{4\pi\ell}{3} (2\mu)^{2/3} \tag{3.17}$$

Thus for large black holes, one finds $\mathcal{A} \simeq 2\ell\mathcal{C}$. It is as if the black hole extends a distance $\ell$ off the brane on each side. In fact, the black hole extends much farther, but its contribution to the area is suppressed by the AdS geometry — see discussion in section 4. The black hole extends to $z_{max} = \ell(1 + 2\mu x_2) \approx (2\mu)^{2/3}\ell$, but this corresponds to a proper distance $L_z \simeq \ell \log(2\mu)$. While $L_z$ is much larger than the AdS scale $\ell$, it is still much smaller than the extent of the horizon on the brane $r = 2\mu\ell$. Hence the large black hole looks like a flattened pancake. This naive discussion can actually be supported by a more general analysis which we will describe in section 4.

### 3.3. Black hole thermodynamics

We would now like to consider the thermodynamic properties of these black holes on the brane. Consider first the Hawking temperature. This is most easily calculated by analytically continuing the time coordinate and fixing the periodicity of the Euclidean section by requiring that the geometry is smooth at the 'horizon' [18]. Since the temperature is constant over the horizon, we can calculate it on the brane, where the metric is (3.8). Given the $(t,r)$ part of this metric is identical to the standard four-dimensional Schwarzschild geometry, we may immediately write

$$T = \frac{1}{8\pi\mu\ell} \ . \tag{3.18}$$

While this result looks like the standard four-dimensional result, it differs since we will see that the actual mass is not proportional to $\mu$.

Next we have the black hole entropy which is determined by the four-dimensional area

$$S = \frac{\mathcal{A}}{4G_4} = \frac{\ell^2\Delta\varphi}{2G_4}\int_0^{x_2}\frac{dx}{(x-y_0)^2} = \frac{4\pi\mu^2\ell^2}{G_4}\frac{1}{(1+2\mu x_2)(1+3\mu x_2)} \tag{3.19}$$

where again $x_2$ is the positive root of $G(x)$ in eq. (3.2). For small $\mu$, $x_2 \simeq 1$ as in eq. (3.10) and $\mu\ell = G_4 M_4$, and so we recover the result for a Schwarzschild black hole with $S \simeq 4\pi G_4 M_4^2$. On the other hand in the large $\mu$ regime, one finds $S = 2\pi\ell^2(2\mu)^{2/3}/(3G_4)$ from eq. (3.16).

Using the first law of black hole thermodynamics

$$\delta M = T\,\delta S \ , \tag{3.20}$$

we can define a four-dimensional thermodynamic mass to the black hole[6]. That is, we need only integrate this differential relation up from zero mass and area to finite values. This may seem a difficult task, given that the entropy (3.19) depends on $x_2$, the root of cubic equation. However, the problem becomes simpler when we introduce the auxiliary variable $\hat{x} = 2\mu x_2$ which satisfies

$$\hat{x}^2(1+\hat{x}) = 4\mu^2 \ . \tag{3.21}$$

---

[6] Since our solutions are not asymptotically AdS in the usual sense, one cannot define the black hole mass from the standard asymptotic formulas.

In terms of this variable the entropy (3.19) and the temperature become

$$S = \frac{\pi \ell^2}{G_4} \frac{\hat{x}^2}{1 + \frac{3}{2}\hat{x}} \ , \qquad T = \frac{1}{4\pi\ell} \frac{1}{\hat{x}\sqrt{1 + \hat{x}}} \ . \tag{3.22}$$

It is now relatively straightforward to integrate the first law (3.20) with the final result that

$$G_4 M_4 = \frac{\ell}{2} \left( 1 - \frac{\sqrt{1 + \hat{x}}}{1 + \frac{3}{2}\hat{x}} \right) \ . \tag{3.23}$$

One can readily verify that this mass formula is monotonic in $\hat{x}$. For small $\mu$ (and hence small $\hat{x} \simeq 2\mu$), one recovers

$$G_4 M_4 \simeq \frac{\hat{x}\ell}{2} \simeq \mu\ell \ . \tag{3.24}$$

Perhaps more surprising is that for large $\mu$ (and hence large $\hat{x} \simeq (2\mu)^{2/3}$), one finds

$$G_4 M_4 \simeq \frac{\ell}{2} \left( 1 - \frac{2}{3(2\mu)^{1/3}} \right) \ . \tag{3.25}$$

Hence for large $\mu$, which corresponds to black holes with arbitrarily large horizon area, there is a limiting mass $G_4 M_{4,max} = \ell/2$. Using (3.13), this agrees exactly with the limiting mass in three dimensions $G_3 M_{3,max} = 1/4$ coming from the fact that the deficit angle cannot exceed $2\pi$:

$$M_{3,max} = M_{4,max} \ . \tag{3.26}$$

One is then emboldened to compare the generic mass formula for three dimensions to eq. (3.23). Essentially, one must express the mass in terms of the auxiliary variable $\hat{x}$ which yields

$$\begin{aligned} G_3 M_3 &= \frac{1}{8\pi} \left( 2\pi - \Delta\varphi \right) = \frac{1}{4} - \frac{1}{2|G'(x_2)|} \\ &= \frac{1}{4} \left( 1 - \frac{\sqrt{1 + \hat{x}}}{1 + \frac{3}{2}\hat{x}} \right) \ . \end{aligned} \tag{3.27}$$

Again using eq. (3.13), one finds that the three-dimensional mass calculated in terms of the deficit angle on the asymptotic brane geometry corresponds precisely to the four-dimensional mass calculated using the relations of black hole mechanics.

Note that a physicist confined to the brane would see a horizon at $r = 2\mu\ell$ with a proper circumference $\mathcal{C} = 2\mu\ell\Delta\varphi$, where for general $\mu$, $\Delta\varphi = 4\pi\mu/[\hat{x}(1 + 3/2\,\hat{x})]$. Hence she may be tempted to ascribe an entropy of

$$S_3 = \frac{\mathcal{C}}{4G_3} = \frac{\pi\ell}{2G_3} \frac{\hat{x}(1 + \hat{x})}{1 + \frac{3}{2}\hat{x}} \ . \tag{3.28}$$

Using eq. (3.13) to compare this expression with that in eq. (3.22), we find that $S_3$ yields approximately the correct result for large $\hat{x}$, that is large $\mu$. This remarkable agreement is because in this regime the black hole horizons are essentially flattened pancakes extending roughly a distance $\ell$ off of the brane. The fact that the black hole entropy is not exactly $\mathcal{C}/4G_3$ should not be a surprise given that there are no black holes in pure $2+1$ gravity without a cosmological constant. The presence of the horizon is already an indication of four-dimensional physics. Since the three- and four-dimensional masses coincide as well as the Hawking temperatures on the brane and in the bulk, a 'flat-world' observer could compute the black hole entropy by integrating the first law. Hence she could infer information about the four-dimensional black hole horizon area from these three-dimensional observations.

## 4. Discussion

We have discussed exact four-dimensional solutions of Einstein's equation with negative cosmological constant describing localized black holes bound to a two-brane. At large distances along the brane, one recovers three-dimensional gravity. This confirms ideas of Randall and Sundrum in a lower dimensional context, where explicit solutions are available. Intuitively, one can understand the origin of lower dimensional gravity in this case as follows. A black hole or any mass on the Planck brane is accelerating, and before the brane is introduced, one needs a mechanism to support this acceleration. In the original AdS C-metric, this force is supplied by a semi-infinite cosmic string. At large transverse distance from this string, the gravitational field is described by $2+1$ gravity. The procedure of removing the cosmic string and introducing the brane does not change the solution off the brane. By continuity, the metric on the brane asymptotically will still be described by $2+1$ gravity. In other words, the boundary conditions on the brane are the same as the ones coming from a semi-infinite cosmic string.

Our results confirm that the singularities that arise when considering just the zero mode in the Randall-Sundrum scenario are physically irrelevant. Localized sources generate localized gravitational fields which fall off near the AdS horizon. In fact, since the AdS horizon is infinitely far away in spatial directions, the field due to sources on the brane is strictly zero at the AdS horizon – the horizon geometry is unchanged.

Even though we recover $2+1$ gravity asymptotically, there can be deviations at rather large scales. In fact, even though there are no black holes in pure $2+1$ gravity, the metric on the brane describes a black hole with a horizon size that can be arbitrarily large. As we have seen, there is a maximum possible mass in $2 + 1$ gravity, and as one approaches this maximum value, the horizon size diverges. If one starts with a distribution of matter on the brane with mass close to the maximum value, it will form a black hole at very low density. As we have discussed, it is also true that arbitrarily large four-dimensional black holes can be produced with finite total energy. It is possible that these solutions have implications that go beyond the Randall-Sundrum scenario.

Throughout this paper we have been considering a model with single (positive tension) brane with an infinite transverse direction [1]. There is another related model [4] where two membranes are introduced, one with positive tension (the "invisible" brane) and another one with negative tension, where the "visible" world is supposed to reside. The transverse dimension is compact in this model, as it is bounded by these two membranes. We want now to investigate what the effect of a black hole localized on the "visible" brane is. To model this situation, we can slice our solution (3.1) at $x = 0$, but this time we discard the region $x > 0$ instead of $x < 0$ as we did before. Then we paste it to a copy of itself along $x = 0$, so the resulting brane has negative tension (for the purpose of this discussion we do not need to worry about the "invisible" brane). However, if we now want to eliminate the conical singularity in the resulting spacetime we have to fix $\Delta\varphi$ to ensure the geometry is smooth on the axis $x = x_1$, instead of $x = x_2$ (at least for small $\mu$, when $x_1$ is real). This produces a conical excess asymptotically on the brane, *i.e.*, a *negative* mass on the brane! Moreover this mass approaches $-\infty$ as the black string regime is reached. Therefore, this two-membrane model produces negative mass black holes on the brane out of four-dimensional black holes with positive mass. In our construction, this is due to the negative tension of the cosmic string that is now pushing on the black hole in the original AdS C-metric (3.1) before the brane is introduced. This is an explicit realization of claims that these models produce anti-gravity on the "visible" brane [19]. Note that for large masses (*i.e.*, $\mu > 1/3\sqrt{3}$), the root $x_1$ disappears, and we would have a black string extending out in the $x < 0$ region. In this case, whether the total solution had an angular deficit or excess would depend heavily on the boundary conditions imposed by the positive tension brane.

At this point the most important question is what are the implications of our results to the more realistic five-dimensional context. It seems likely that the gravitational field of a black hole on a three-brane will again fall off far from the brane to leave the AdS horizon nonsingular[7] (and, in fact, unchanged). Recall that the effects of the black hole fall off like $1/z$ far from the Planck two-brane. Extending this result to $n+1$ dimensions, we expect that approaching the horizon at fixed radius, the fall-off of the metric to be identical to an $(n + 1)$-dimensional black hole, *i.e.,* $G_{n+1}M/z^{n-2}$ for large $z$. This then is the additional fall off beyond the overall factor of $\ell^2/z^2$ in the AdS metric. The latter corresponds to the profile of a zero-mode excitation, as discussed in section 2. Hence thinking in terms of a mode expansion, the Kaluza-Klein modes effectively smooth out the zero-mode contribution, to produce the more dramatic decay of the gravitational field and to avoid the appearance of any singularity at the horizon.

A higher dimensional black hole will also modify the asymptotic gravitational field on the brane. The results for our four-dimensional solutions suggest that a five-dimensional black hole will introduce $\ell G_4 M/r^2$ corrections to the $G_4 M/r$ gravitational potential on the three-brane. Note that this differs from the $\ell^2 G_4 M/r^3$ corrections predicted in [1]. These corrections would still be negligible for solar mass black holes. Similarly in the $(n+1)$-dimensional construction for gravity on an $(n–1)$-brane in section 2, the leading potential on the brane is $G_n M/r^{n-3}$, and the leading correction would be proportional to $\ell G_n M/r^{n-2}$. Hence these still become insignificant for distances $r > \ell$ on the brane. In our lower dimensional example, we found that significant deviations extended out to $r \sim \mu\ell$, which can be significantly larger than the AdS scale $\ell$ alone. Again this is an artifact of 2+1 dimensions, where there are no deviations from flatness to compete with the deviations arising from the higher dimensions.

It is interesting to compare the corrections to the Newtonian potential to those which might arise due to higher curvature interactions. Let us consider the simplest case of adding a curvature-squared interaction to Einstein gravity in $n$ dimensions, *i.e.,* consider the effective action

$$I = \frac{1}{16\pi G_n} \int d^n x \, \left( \mathcal{R} + \ell_c^2 \, \mathcal{R}_{\mu\nu r\sigma} \mathcal{R}^{\mu\nu r\sigma} \right) \qquad (4.1)$$

where the length $\ell_c$ would be proportional to some cut-off scale. To leading order, the gravitational potential corresponds to that of Einstein gravity with $h_{\mu\nu} \propto G_n M/r^{n-3}$, which then induces curvatures of the order $\mathcal{R}_{\mu\nu r\sigma} \propto G_n M/r^{n-1}$ (both in an appropriate asymptotically flat frame). Schematically then these curvatures introduce a new source term in the modified field equations of the form: $\partial^2 h' \simeq \ell_c^2 \, \partial^2 \mathcal{R}$. Hence one expects the corrections to the gravitational potential to be of the order $h'_{\mu\nu} \propto \ell_c^2 G_n M/r^{n-1}$. We

---

remark that this result agrees with those from an explicit detailed analysis, *e.g.*, [21]. Hence we see these corrections are suppressed by $\ell_c^2/r^2$ relative to the leading Newtonian terms, which is higher order compared to the $\ell/r$ suppression appearing in the Randall-Sundrum geometry.

Actually, for the case of $n = 4$, the suppression of contributions arising from higher curvature interactions is even stronger than the schematic analysis above indicates. We did not consider interactions involving the Ricci tensor or scalar, *i.e.*, $\mathcal{R}_{\mu\nu}\mathcal{R}^{\mu\nu}$ or $\mathcal{R}^2$, since their contributions vanish by the leading order equations of motion. Hence in the effective action (4.1), the curvature squared interaction is equivalent to the four-dimensional Euler density which would include additional Ricci squared terms. However, since the later is a topological term in four dimensions, it will not affect the equations of motion for $n = 4$.[8] Hence a detailed analysis [21] would show that the correction $\ell_c^2 G_n M/r^{n-1}$ appearing in the schematic analysis comes with a coefficient of precisely zero for $n = 4$. Hence to produce a nonvanishing result in four dimensions, one would have to consider a curvature cubed interaction which would then give rise to a correction $h'_{\mu\nu} \simeq \ell_c^4 (G_4 M)^2/r^6$. In any event, this is greatly suppressed compared to the $\ell G_4 M/r^2$ correction that appears in the Randall-Sundrum scenario.

We might comment that experimental tests have already probed the $1/r^2$ corrections to the Newtonian gravitational potential — see *e.g.*, [23]. These observations are in good agreement with Einstein gravity, which by itself predicts post-Newtonian corrections of the form $(G_4 M)^2/r^2$ because of the nonlinearities of the gravitational field equations. Of course, in the context of solar system tests or any astrophysical context, we can expect that these Einstein corrections will swamp the brane-world corrections since $G_4 M \gg \ell$.

Just as in section 3.3, one can expect that the thermodynamic properties of $(n+1)$-dimensional black holes bound to the Planck brane will closely approximate those of black holes arising in $n$-dimensional Einstein gravity, at least for large black holes. In the latter case, the black hole geometry on the brane will closely resemble that of an Einstein solution in $n$-dimensions, and so the surface gravity or Hawking temperature will be very close to the expected value. The corrections will be suppressed by $\ell (G_n M)^{-1/(n-3)}$. One of the striking results in this section was that the four-dimensional thermodynamic mass agreed precisely with the standard asymptotic mass calculated on the brane. It would be interesting to see

---

[8] This is precisely the same cancellation which renders pure Einstein gravity in four dimensions finite at one-loop [22].

if this result extends to higher dimensions.

In section 3.3, we found that the naive three-dimensional entropy $\mathcal{C}/4G_3$ matched quite well with $S = \mathcal{A}/4G_4$ for large black holes — all the more surprising since we did not have a three-dimensional theory of gravity with black holes. In fact, this remarkable agreement is expected to hold in arbitrary dimensions. The point is that most of the area on the higher dimensional event horizon comes from a region very near to the Planck brane. Consider approximating the area of a large $(n{+}1)$-dimensional black hole as a 'cylinder' which extends from the brane at $z = \ell$ to $z_{max}$ with fixed radius $r$, and then closes off with a 'disk' with fixed $z = z_{max}$. First, note that the 'circumference' of the horizon measured on the brane is

$$\mathcal{C} = r^{n-2}\Omega_{n-2} \tag{4.2}$$

where $\Omega_{n-2}$ is the area of a unit $(n{-}2)$-sphere. Now, the contribution to the $(n{+}1)$-dimensional area from the cylinder is

$$\begin{aligned}
\mathcal{A}_c &= 2r^{n-2}\Omega_{n-2} \int_\ell^{z_{max}} \left(\frac{\ell}{z}\right)^{n-1} dz \\
&= \frac{2\ell}{n-2}r^{n-2}\Omega_{n-2}\left[1 - (\ell/z_{max})^{n-2}\right]
\end{aligned} \tag{4.3}$$

while the contribution of the disk is

$$\mathcal{A}_d = \frac{2}{n-2}r^{n-1}\Omega_{n-2}\left(\ell/z_{max}\right)^{n-1} \ . \tag{4.4}$$

Hence if the horizon extends into the bulk such that $\ell/z_{max} \ll 1$, one has to leading order that

$$\mathcal{A} = \frac{2\ell}{n-2}\mathcal{C} \ . \tag{4.5}$$

Now further using the relation (2.8) between the Newton's constants on the brane and in the bulk, one has

$$S = \frac{\mathcal{A}}{4G_{n+1}} = \frac{\mathcal{C}}{4G_n} \ . \tag{4.6}$$

Hence we see that the AdS geometry conspires to reproduce precisely the black hole entropy of $n$-dimensional Einstein gravity.

Of course, one does not expect the correspondence (4.6) of the bulk and brane entropies will continue to hold for small black holes. As indicated in eqs. (4.3) and (4.4), there are corrections to the relation (4.5), which are suppressed by $\ell/z_{max}$, revealing the higher dimensional nature of the black hole. Of course, the crude model above will not give the detailed form of these corrections. In passing, we note that for black strings constructed by letting $g_{\mu\nu}$ in eq. (2.1) be an $n$-dimensional black hole, as in [24], these corrections vanish since $z_{max} \to \infty$. Hence the correspondence (4.6) of the black hole entropy calculated in the bulk and on the Planck brane would be exact.

On quite general grounds one expects the surface gravity to be constant over the event horizon [25], and so the Hawking temperature will be the same on the brane and in the bulk spacetime. However, it appears that large black holes will prefer to radiate in the brane. This is not just because of the large number of species of particles on the brane compared to just gravity in the bulk. One can crudely estimate the decay rate as $dM/dt \sim \mathcal{A}T^d$, where $\mathcal{A}$ is the horizon area and $d$ is the relevent spacetime dimension. Hence, for example, for a large Schwarzschild black hole of radius $r_0$ on the three-brane, with $d = 4$ and $T \sim 1/r_0$, we have the usual result $dM/dt \sim 1/r_0^2$. From the five-dimensional standpoint, $d = 5$ and the area will be approximately $\ell r_0^2$ — as above. Thus gravitational Hawking radiation into the bulk would contribute $dM/dt \sim \ell/r_0^3$, which will be much smaller. Small black holes will behave differently. When $G_4 M < \ell$, the black hole behaves like a five-dimensional Schwarzschild solution. If the radius of the event horizon is $r_0$, one still has $T \sim 1/r_0$. Hence the four-dimensional radiation still yields $dM/dt \sim 1/r_0^2$. The five-dimensional area is now approximately $r_0^3$, and so the bulk radiation also contributes $dM/dt \sim 1/r_0^2$. However in this case, the latter contribution is smaller because the black hole only radiates into the bulk with a single species, *i.e.*, the five-dimensional graviton, while the brane radiation involves all of the species of brane-matter fields. Still, for these small black holes, one should expect that the relation between $r_0$ and $2G_4 M$ is modified, and so the evolution of these small black holes would in principle be distinguishable from that those in pure Einstein gravity. This may then have implications for primordial black holes [26].

In section 3.2, we found that the geometry of large black holes resembled a flattened pancake. Note that on purely dimensional grounds, one can expect that for large $G_n M$

$$z_{max} \simeq \ell \left( G_n M / \ell^{n-3} \right)^{\alpha} \tag{4.7}$$

for some constant $\alpha$. From this observation (independent of the value of $\alpha$), one finds as a robust result that the proper distance that these black holes extend off the Planck brane is

$$L_z \simeq \ell \log \left( G_n M / \ell^{n-3} \right) . \tag{4.8}$$

Hence one can expect quite generally that for large black holes, $L_z$ is larger than the AdS scale $\ell$, but still much smaller than the extent of the horizon on the brane $r \sim (G_n M)^{1/(n-3)}$. Hence these large black holes will always have a pancake geometry.

The authors of [24] presented the interesting idea that the physical extent of the black holes off of the Planck brane is related to the Gregory-Laflamme instability [27]. Suppose we start with the black string described by (2.1) where $g_{\mu\nu}$ is the metric of a black hole of mass $M$. Since the proper distances on surfaces of constant $z$ are shrinking as $\tilde{r} = \ell r/z$ due to the AdS geometry, we have an effective mass scale at finite $z$ of

$$(G_n M)^* = (\ell/z)^{n-3} G_n M \ . \tag{4.9}$$

The idea in [24] is that the horizon then can not extend far past the limit where $(G_n M)^* \simeq \ell^{n-3}$, otherwise it becomes long and narrow, and the Gregory-Laflamme instability would arise due to long-wavelength metric fluctuations. The cut-off arises because the confining nature of the AdS geometry only allows wavelengths shorter than $\ell$. If we assume that the instability sets in right at $(G_n M)^* \simeq \ell^{n-3}$, we find

$$z_{max} = \ell \left( G_n M / \ell^{n-3} \right)^{1/(n-3)} \tag{4.10}$$

and then

$$L_z \simeq \frac{\ell}{n-3} \log \left( G_n M / \ell^{n-3} \right) \ . \tag{4.11}$$

Hence this suggestion [24] yields a result consistent with (4.8), and again predicts a pancake-like geometry for large black holes on the Planck brane (see also [20]).

To construct an exact solution describing a black hole bound to a three-brane, one would need to start with a five-dimensional analog of the C-metric. Unfortunately, such a metric is not yet known. However, the following qualitative features must be present. To accelerate the black hole, one needs an analog of the cosmic string. In four spatial dimensions, if one compresses a string of matter to make it thinner and thinner, it will eventually form a horizon and become a black string. Thus we expect a vacuum solution describing an accelerating five-dimensional black hole to actually contain a black string. If the string were too thin, it would break up into individual black holes [27]. So the thickness of the black string must be at least comparable to the AdS scale. This is similar to the picture arrived at in [24]. Once one has the accelerating black string solution, one should be able to cut off the asymptotic region and introduce a three-brane as we did in lower dimensions.

## Acknowledgements

GTH and RCM would like to thank the participants of the ITP Program on Supersymmetric Gauge Dynamics and String Theory for stimulating discussions, especially Steve Giddings, Lisa Randall and Raman Sundrum. The work of RE is partially supported by UPV grant 063.310-EB187/98. GTH was supported in part by NSF Grant PHY95-07065. RCM was supported in part by NSERC of Canada and Fonds du Québec. At the ITP, RCM was supported by PHY94-07194. This paper has report numbers: EHU-FT/9914, NSF-ITP-99-134, and McGill/99-35.